# Coherent plasmonic absorption in the femtosecond regime

Venkatram Nalla[1,*], Xu Fang[2], João Valente[2], Handong Sun[1], and Nikolay I Zheludev[1, 2]

*[1]Centre for Disruptive Photonic Technologies, The Photonic Institute,
School of Physical and Mathematical Sciences, Nanyang Technological University, Singapore*

*[2]Optoelectronics Research Centre and Centre for Photonic Metamaterials,
University of Southampton, Highfield, Southampton, UK*

Dissipation of electromagnetic energy through absorption is a fundamental process that underpins phenomena ranging from photovoltaics to photography, analytical spectroscopy, photosynthesis, and human vision. Absorption is also a dynamic process that depends on the duration of the optical illumination. Here, we report on the resonant plasmonic absorption of a nanostructured metamaterial and the non-resonant absorption of an unstructured gold film at different optical pulse durations. By examining the absorption in travelling and standing waves, we observe a plasmonic relaxation time of 11 fs as the characteristic transition time. The metamaterial acts as a beam-splitter with low absorption for shorter pulses, while as a good absorber for longer pulses. The transient nature of the absorption puts a frequency limit of ~90 THz on the bandwidth of coherently-controlled, all-optical switching devices, which is still a thousand times faster than other leading switching technologies. © Anita Publications. All rights reserved.

## 1 Introduction

Light absorption is a fundamental optical process, controlling phenomena at scales from the planetary (e.g. climate on the Earth regulated by the strength of sunshine) to the microscopic (e.g. the physiological process of human vision). It underpins or strongly affects the functionalities of almost all modern, light-enabled technologies. Strong light absorption can be highly desirable [1,2]; it can also be a foe, for instance in optical telecommunication networks [3]. The ability of a material to absorb light is widely seen as an inherent and unchangeable property of the material. However, it recently came to the attention of research communities that in bulk media [4] and thin films, [5-9] absorption manifests itself differently in travelling and standing waves, allowing for a transition from complete power dissipation (i.e. coherent perfect absorption) to perfect transparency in a thin absorber (i.e. coherent perfect transmission). This transition, termed as the coherent control of absorption, can be achieved by changing the mutual phase or the intensity balance of two coherent counter-propagating waves interacting on the absorber. For instance, a thin film absorbing 50% of light from a travelling wave can have any value of absorption from 0 to 100% in a standing wave. Such ability to control energy dissipation by interfering waves on a thin absorbing layer provides exciting new opportunities for energy harvesting, radiation detection, dark pulse generation, all-optical data and image processing [10-16].

As compared to absolute value of the absorption, the temporal dynamics of absorption is of equal, sometimes even more, important for many applications [17-19]. In the regime of linear absorption, the

*Corresponding author*
*e mail: nallavram@gmail.com* (Venkatram Nalla)

intensity of incident light is insufficient to initiate nonlinear, multi-photon processes. It is often reasonable to assume that the energy dissipation does not depend on the temporal profile of the excitation light. However, demanding technological applications, such as the emerging all-optical data and image processing enabled by coherently-controlled absorption [13,20-23], aim to exploit high-bandwidth operation by using extremely short optical pulses of only a few oscillations. In this regime, the assumption that light absorption is independent of pulse duration can be challenged and must be tested. Our particular interest is in the transient characteristics of the absorption of nanostructured plasmonic films in both standing and travelling waves. Indeed, such films were recently used to demonstrate various applications of perfect absorption across a wide range of research fields [5-7,20]. In such absorbers, the energy dissipation occurs through the coupling between light and the plasmonic modes of the nanostructure, which is channelled through radiative and non-radiative processes. How fast does the light absorption in standing and travelling waves develop? What determines the dynamics of the absorption and subsequently the bandwidth of absorption-based technologies? In this work, we aim to answer these questions, and will assess the ultimate speed at which the coherent absorption process occurs. By using optical pulses with duration from 6 fs to 185 fs, we show that the characteristic time for absorption to settle in such materials is ~11 fs. This value originates from the plasmon relaxation time and sets a maximal bandwidth of ~90 THz for, coherent absorption-based, all-optical switching.

Although the microscopic mechanism of energy dissipation in travelling and standing waves is believed to be identical, we decided to measure light absorption in both waves. Indeed, in a travelling wave experiment, energy loss can be easily evaluated from transmission and reflection. In comparison, a standing wave experiment, although significantly more complicated, can be configured to show variable energy loss in the sample. Moreover, a standing wave experiment allows probing absorption in the same configuration used for the coherent control of light with light, therefore providing a direct way of assessing the bandwidth of all-optical coherent switching. We are not aware or any other work that compares the transient dynamics of linear plasmonic absorption between standing and travelling waves.

## 2 Methods

*Measurement of ultrafast standing wave absorption*

Laser pulses with duration of 6 fs were generated by a mode-locked Ti:sapphire laser (Femtolasers, Rainbow). They were sent into a spatial light modulator (Biophotonics, MIIPS Pulse Shaper) to tune both the central wavelength and the spectral width. The spectral width ranged from 5 nm to 165 nm, corresponding to a pulse duration from 185 fs to 6 fs. The pulse spectrum was measured using a USB spectrometer (Ocean optics, USB 4000), and the pulse duration was measured using a homebuilt autocorrelator. The product of the frequency bandwidth and the pulse duration was ~0.44, indicating a transform-limited Gaussian profile [24]. .A homebuilt interferometer similar to [5,7] was used to measure the coherent absorption of the samples, with its schematic illustrated in Fig 1. The laser beam was split into two identical pulses ($\alpha$ and $\beta$) using a 50:50 pellicle beam-splitter. Two focusing mirrors (ROC-15 cm) were used to focus them onto the sample. A piezoelectric translation stage was used to change the position of the sample. The average power at the sample position was balanced using a variable neutral density filter for each wavelength and pulse duration. It was maintained at a low level to suppress thermal damage and nonlinear effects. Pulse dispersion induced by optical components was compensated by giving feedback to the pulse shaper. A mechanical chopper was used in one of the arms to increase the signal to noise ratio. The two output beams $\gamma$ and $\delta$ (transmitted and reflected from both sides of the sample) were directed to a pair of identical photodiodes, the outputs from which are monitored using lock-in amplifiers referenced to the chopping frequency of 130 Hz. The output of the photodiodes was calibrated for each measurement using an open window on the sample, in order to eliminate the undesired influence of optical components on the result (e.g. variation in the power splitting ratio of the pellicle with wavelength).

*Metamaterial fabrication*

The metamaterial sample consisted of a 50 μm × 50 μm array of asymmetric split ring resonators, which was milled on a free-standing gold film 60 nm in thickness. The sample showed a plasmonic resonance at ~800 nm, where the conventional, single-beam absorption was close to 47%. The sample fabrication involved three main steps. Thermal evaporation at low pressure (2x10$^{-6}$ mbar) and low deposition rate ($\approx$ 0.10 nm/s) was used to deposit a 60 nm thick gold film on a commercially available 50 nm thick SiN$_x$ membrane. Reactive ion etching (Plasma Lab 80 plus, Oxford Instruments) was then applied (CHF$_3$ 2 sccm, Ar 10 sccm, pressure 50 mTorr, power 80 W, and time 10 minutes) to the backside of the sample to completely remove the SiN$_x$ and expose the gold layer. Finally, focused ion beam milling (FEI, Helios 600 NanoLab) was used to fabricate the metamaterial nanostructures. Ar 10 sccm, pressure 50 mTorr, power 80 W, and time 10 minutes) to the backside of the sample to completely remove the SiN$_x$ and expose the gold layer. Finally, focused ion beam milling (FEI, Helios 600 NanoLab) was used to fabricate the metamaterial nanostructures.

## 3 Results

*Experimental setup for coherent absorption measurement*

Figure 1. shows the schematic of the experimental setup for measuring the coherent, standing-wave absorption and the conventional, travelling-wave absorption (see Methods for details). A Ti:sapphire femtosecond laser (pulse duration ~6 fs, central wavelength tuneable around 800 nm, repetition rate 75 MHz) was used as the light source in both the measurements. Output pulses from the laser were spectrally filtered and conditioned by a pulse shaper to tune their spectral and temporal profiles. For all the pulses, the product of the pulse frequency bandwidth ($\Delta\nu$) and the pulse duration ($\Delta\tau$) was measured to be close to $\Delta\nu \times \Delta\tau \approx 0.44$ (Fig 1A, inset), which indicated a transform-limited Gaussian profile [24]. The pulses were subsequently directed into an interferometer built by using a 50:50 pellicle beam-splitter and two broadband focusing metal mirrors. The mirrors, which had a long ROC of 15 cm, focused light onto the sample from both sides, forming a standing wave at its position. The sample was placed on a piezoelectric translation stage. The output light was collected using a pair of conventional silicon detectors. We measured two samples, a metamaterial plasmonic absorber and an unstructured gold thin film. The metamaterial was a two-dimensional square array of asymmetric split-ring metamolecules fabricated on a gold film (Fig 1B). Both the samples were on the same free-standing film with thickness 60 nm.

*Travelling wave absorption*

The spectra of both samples were first measured using a conventional microspectrometer (Fig S1). They reveal the static properties of the samples under travelling wave illumination. Corresponding dynamic properties of the samples were subsequently investigated using modified interferometer setup, by blocking one of the incident beams and changing the pulse duration.

Figure 2 shows the absorption, reflection, and transmission of both samples under the illumination of pulses with different durations. The central wavelength is 800 nm, overlapping with the peak wavelength of the plasmonic absorption of the metamaterial. The transmission and reflection of the metamaterial show very similar characteristics: they rapidly fall from ~32% at the pulse duration of 6 fs to ~26% at 11 fs, and stabilise at this level for longer pulses (Fig 2A). In comparison, the absorption grows from ~36% at 6 fs to ~47% at 11 fs, and stabilises at this level for longer pulses. For the unstructured gold film (Fig 2B), the spectra show a much smaller, smoother change with pulse duration. In the whole measurement range from 6 fs to 185 fs, they vary by less than 3% in absolute value.

*Standing wave absorption*

In the standing wave absorption experiments, the absorber acts as a "4-port" device with two input ports $\alpha$ and $\beta$ (for incident beams) and two output ports $\gamma$ and $\delta$ (for outgoing beams), as annotated in Fig 1A.

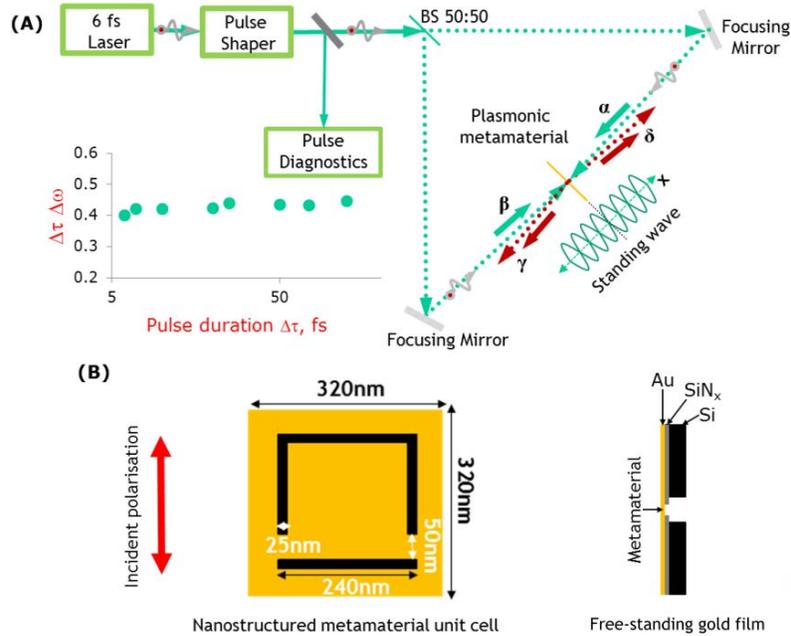

 **Characterising light absorption in a standing wave and a traveling wave regimes.** A) An ultra-thin sample is placed in a standing wave formed by two counter-propagating femtosecond pulses. The standing wave absorption is evaluated by comparing the output in channels γ and δ with the input in channels α and β. The travelling wave absorption is evaluated by blocking channel β. B) The unit cell of the metamaterial sample, which was fabricated on an ultra-thin, free-standing gold film.

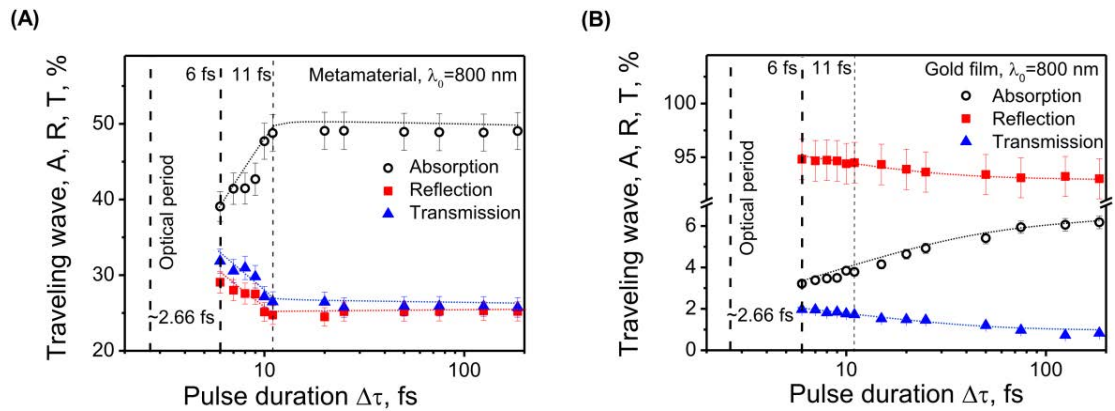

Fig 2. Interaction between thin absorbers and short pulses in a traveling wave. Dependence of absorption, reflection and transmission on pulse duration Δτ for (A) the plasmonic metamaterial and (B) the unstructured gold film. The central wavelength of the pulse is 800 nm for all the pulse durations.

The standing wave absorption is a function of the sample position $x$ in the standing wave. It is obtained by comparing the output intensities, $I_\gamma(x)$ and $I_\delta(x)$, with the input intensities $I_\alpha$ and $I_\beta$. $I_\alpha$ and $I_\beta$ are equal, and their sum $I_0 = I_\alpha + I_\beta$ is at a very low level (total pulse fluence less than 0.5 nJ/cm²). Figure 3 shows the normalized output (output/input) for the pulse duration Δτ = 6 fs and the central wavelength $\lambda_0 = 800$ nm. At the central antinode of the standing wave, absorption is the highest and the total output $I_\gamma + I_\delta$ is nearly at the minimum. Displacing the film away from this position causes oscillation in the output, with its local maximum corresponding to the nodes of the standing wave. Due to the limited temporal coherence of the pulses, the oscillation fades at large displacement. The intensity oscillation in channels $\gamma$ and $\delta$ also shows a noticeable phase shift $\Delta\varphi_{\gamma\delta}$ (Fig 3B), the nature of which will be discussed below. $I_{\gamma}(x)$ and $I_{\delta}(x)$ are also slightly different in the amplitude, which is due to the small physical asymmetry of the sample's two sides (see Method for more details). The coherent, standing-wave absorption is evaluated as $A_s(x) = 1 - (I_{\gamma}(x) + I_{\delta}(x))/I_0$. $A_s(x)$ reaches the maximum value $A_s(0)$ at $x = 0$, which we simply denote as $A_s$.

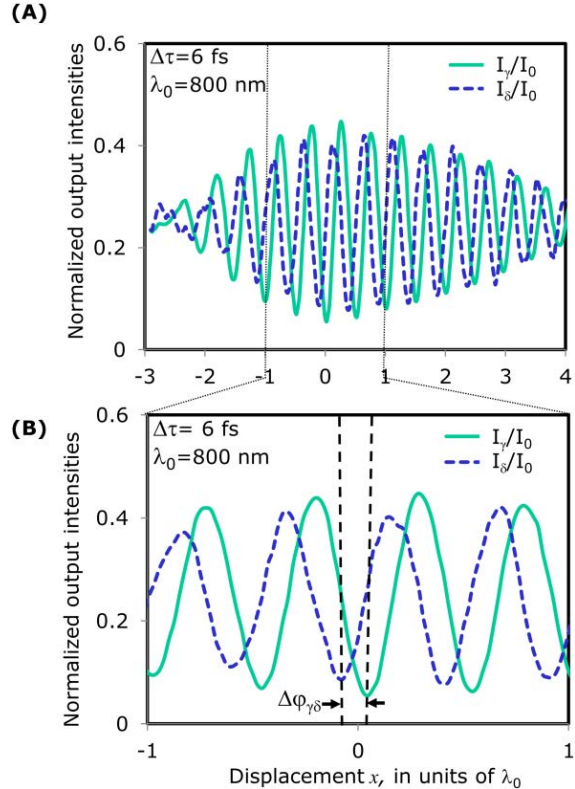

Fig 3. **Metamaterial absorption in an ultrashort standing wave.** (A) Dependence of the normalised output intensities $I_{\gamma}(x)/I_0$ and $I_{\delta}(x)/I_0$ on the sample position $x$. $x$ is normalised against the central wavelength $\lambda_0$, and its origin is at the central point of the standing wave. (B) Enlarged view of Fig 3A shows the phase shift $\Delta\varphi_{\gamma\delta}$ between the two output channels. The pulse duration $\Delta\tau$ is 6 fs, and its central wavelength $\lambda_0$ is 800 nm.

We then measured the dependence of the maximal standing wave absorption $A_s$ on the optical pulse duration $\Delta\tau$ (Fig 4). The central wavelength of the pulse was kept at 800 nm in changing the pulse durations. Figure 4A shows that, for the plasmonic metamaterial, $A_s$ is low for short pulses. It rapidly increases with pulse duration and reaches its maximum of 91% at $\Delta\tau \approx 11$ fs. Figure 4B shows the corresponding data of the unstructured gold film for comparison. A steady but much weaker increase in $A_s$, from 6.5% at 6 fs to 11% at 185 fs, is observed. Therefore, we can conclude that the metamaterial film becomes a better absorber at longer pulse duration, as witnessed in both the standing and travelling wave absorption measurements. Meanwhile, absorption of the unstructured gold film remains small at all pulse durations. In addition, an interesting difference is observed in the phase shift $\Delta\varphi_{\gamma\delta}$ between the metamaterial and the gold film (Fig 4).

For the plasmonic metamaterial, $\Delta\varphi_{\gamma\delta}$ is high at small pulse durations, and rapidly falls to nearly zero once pulse duration reaches ~11 fs. In contrast, $\Delta\varphi_{\gamma\delta} \approx \pi$ for the gold thin film in the whole measurement range.

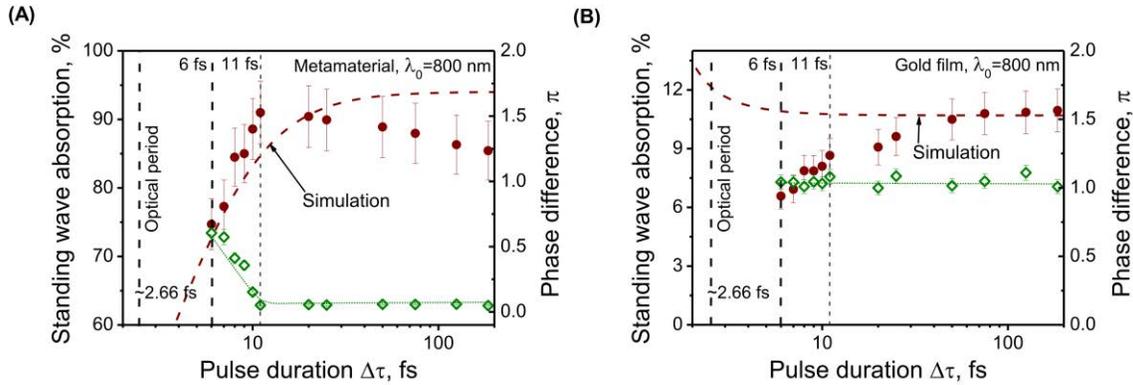

**Fig 4. Dependence of the standing wave absorption $A_s$ and phase shift $\Delta\varphi_{\gamma\delta}$ on the pulse duration $\Delta\tau$.** (A) Both the absorption (brown dots) and the phase difference (green diamonds) are measured for the metamaterial, and the former is reproduced analytically (brown dashed line). Green dashed lines are trending lines. (B) Corresponding data for the unstructured gold film. The pulse central wavelength is 800 nm during all the durations.

To further explore the properties of the maximal standing wave absorption $A_s$, we measured its dependence on the central wavelength $\lambda_0$ of the laser pulses (Fig 5). The pulse duration is fixed at $\Delta\tau \approx 11$ fs. For both samples, the absorption spectra show similar features between the standing and travelling waves. For the metamaterial, the peak value of $A_s$ is at $\lambda_0 = 800$ nm, which coincides with the peak wavelength of the travelling wave absorption.

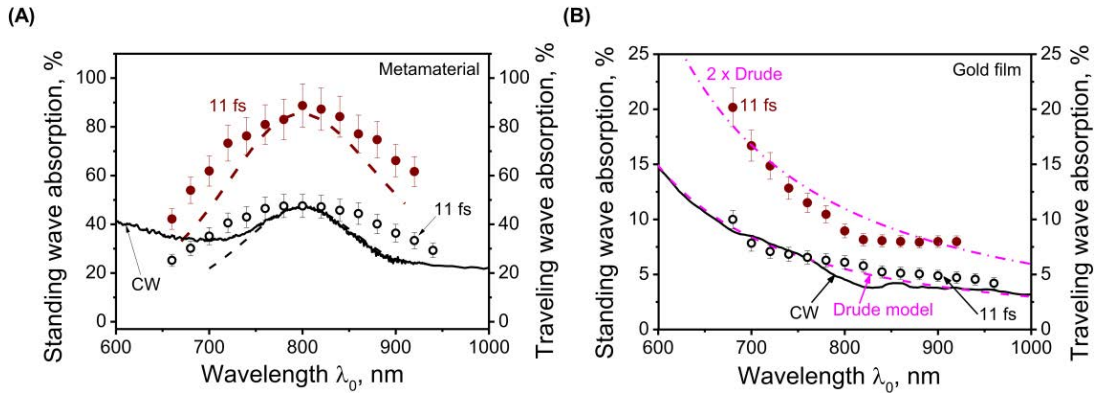

**Fig 5. Spectral dependence of travelling and standing wave absorption.** (A) For the metamaterial, standing (brown dots) and travelling wave absorption (black circles) of ultrashort pulses were measured using the home-built interferometer. The data are overlaid with analytical values (dashed lines) and absorption of c.w. light measured using a conventional spectrometer (black continuous line). (B) Corresponding results for the gold thin film. The theoretical curves (purple lines) are obtained using the Drude model with the actual film thickness taking into consideration.

## 4 Discussion

Figure 4 provides a very interesting insight into the properties of the metamaterial and the unstructured gold thin film, and it is via the phase difference $\Delta\varphi_{\gamma\delta}$. For the gold film, $\Delta\varphi_{\gamma\delta} \sim \pi$ for all the measured pulse durations. Meanwhile, for the metamaterial, $\Delta\varphi_{\gamma\delta} \sim 0$ at long pulse durations, and it quickly increases once the pulse duration decreases below 11 fs. Rather than analysing the complicated temporal variation of the pulses directly, here we first conduct analysis on c.w. light to reveal the key aspects of absorption dynamics.

Figure 6A shows a hypothetical lossless beam-splitter which reflects and transmits a travelling wave in equal proportion. Figure 6B shows an ideal thin absorber that absorbs, reflects, and transmits a travelling wave at proportions of 50:25:25. It may be easily shown (see Supplementary Materials for details) that, the phase difference $\Delta\varphi_{\gamma\delta}$ between the two output beams $I_\gamma$ and $I_\delta$ is out of phase (i.e. $\Delta\varphi_{\gamma\delta} = \pi$) for the former and in phase (i.e. $\Delta\varphi_{\gamma\delta} = 0$) for the latter. This observation holds for any position of the thin film in the standing wave. Therefore, the gold film across the whole range of pulse durations resembles a perfect semi-reflector (i.e. $\Delta\varphi_{\gamma\delta} = \pi$). Meanwhile, the plasmonic absorber behaves likes an ideal absorber at longer pulses (i.e. $\Delta\varphi_{\gamma\delta} = 0$), and becomes closer to a perfect semi-reflector at short pulses.

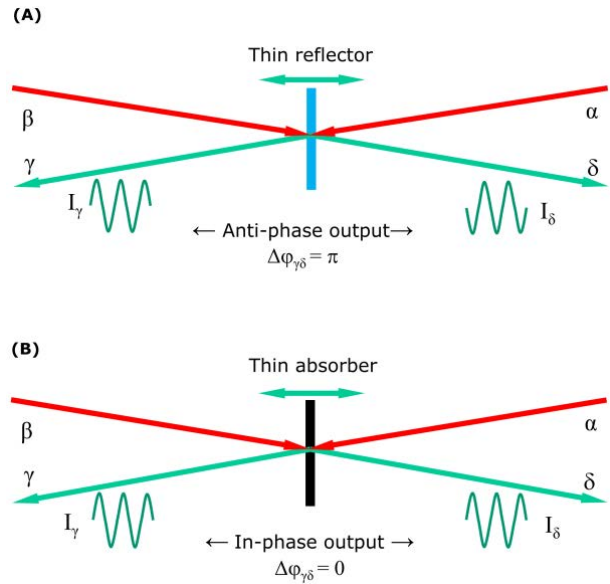

Fig 6. **An ideal lossless 50:50 beam splitter (or semi-reflector) and an ideal thin-film absorber in a standing wave.** Regardless of the position of the thin film along the standing wave, the phase difference $\Delta\varphi_{\gamma\delta}$ between the two output signals $I_\gamma$ and $I_\delta$ is always (A) $\pi$ out of phase for the thin semi-reflector, and (B) in phase (i.e. $\Delta\varphi_{\gamma\delta} = 0$) for the ideal thin absorber.

*Lorentz oscillator model for absorption*

Further analysis is conducted by describing the localized plasmonic resonance as a harmonic oscillator:

$$\frac{d^2y}{dt^2} = \Gamma_p \frac{dy}{dt} + \omega_0^2 y = \frac{-q}{m}E$$

where $y$ is the oscillator displacement, $t$ is the time, $\Gamma_p = 2\pi/\tau_p$ is the localized plasmon damping parameter, $\omega_0$ is the localized plasmon frequency, $-q$ is the electric charge of the oscillator, $m$ is the mass of the oscillator, and $E$ is the driving electric field of the optical wave. Here, $\omega_0 = 2.36\times10^{15}$ Hz, which corresponds to the centre of the plasmonic absorption peak at 800 nm. Here we take $\Gamma_p$ from the measured spectral width of the plasmonic absorption resonance (black dashed line in Fig 5A): $\Gamma_p = 0.59\times10^{15}$ Hz. It gives $\Gamma_p \approx \omega_0/4$ and $\tau_p = 2\pi/\Gamma_p = 10.6$ fs.

The plasmon relaxation has two channels, the first related to absorptive loss (e.g. induced by electron-electron and electron-phonon collisions) and the second to radiation loss. Therefore $1/\tau_p = 1/\tau_c + 1/\tau_r$, where $\tau_c$ is the characteristic electronic collision time in gold and $\tau_r$ is the relaxation time due to radiation to free space. Assuming literature value of the electron collision time $\tau_c = 15$ fs, [25,26] and plasmon relaxation time $\tau_p = 10.6$ fs, we can evaluate radiation loss time $\tau_s = 36$ fs.

A similar equation is applicable to the electron motion in unstructured gold thin film, assuming zero restoring force (i.e. the Drude model):

$$\frac{d^2y}{dt^2} = \Gamma_c \frac{dy}{dt} = \frac{-q}{m}E$$

Here, radiation losses are absent and relaxation only depends on electronic collisions $\Gamma_c = 2\pi/\tau_c$, where $\tau_c = 15$ fs [25,26].

We can readily calculate the travelling and standing wave absorptions of the metamaterial and the unstructured gold. It is the portion $Q_d/Q$ of the optical pulse energy $Q = \int_0^\infty EE^* dt$ that is dissipated by the moving electrons, where

$$Q_d = \int_0^\infty -qE\frac{dy}{dt}\,dt$$

Figure 4 presents the analytically calculated value at different pulse durations, and Fig 5A that at different central wavelengths. The peak absorption is the only free parameter in deriving these results, which show good agreement with the experimental data in both cases.

Although the simple oscillator model well describes all the main trends of traveling and standing wave absorption (i.e. the dependence on pulse duration and its central wavelength), some deviation between the theoretical and experimental curves are inevitable consequences of the simplicity of the model. In Fig 4B the 4% difference in absorption predicted by the model and measured in the experiment for short pulses is related to the finite thickness of the film that is not taken into account in the model. This is confirmed in Fig 5A, where the absorption spectrum is calculated using the Drude model [25,27] with the actual film thickness taken into account.

*The importance of plasmonic resonance*

The small dependence on pulse duration shown in Fig 4B is too weak to reveal any Drude absorption dynamics, which is controlled by electron collisions rate of where $\tau_c = 15$ fs. On the contrary, once structured with asymmetric split ring slits, a gold film becomes a strong plasmonic absorber, and the resonant absorption is significantly influenced by the excitation of plasmonic mode [28]. Indeed, the metamaterial exhibits very different optical properties from the unstructured gold film. Although relatively small for short pulses, the coherent absorption is fully established once the pulse is longer than 11 fs. We argue that, this is the time necessary to build up plasmonic absorption in the sample. Combining with the reflection and transmission spectra measured in the travelling wave, we conclude that: for long pulses, the metamaterial behaves essentially as an ideal thin-film absorber; for short pulses, it behaves like a semi-reflector, or a lossless beam-splitter. The transition pulse duration is determined by the plasmonic resonance. Further analysis is able to show that, making the absorption peak broader (i.e. increasing the plasmonic damping) creates higher coherent absorption for short pulses, which essentially allows for higher operation speed of the metamaterial as a coherent switch (Fig S2).

## 5 Conclusions

In summary, we have analysed the influence of surface plasmons on ultrafast coherent absorption. Standing wave absorption is as high as ~91% for a plasmonic metamaterial, comparing with that of only

~11% for an unstructured gold film, clearly shows the crucial role of plasmons for the coherent absorption. As revealed by interrogating the sample with various pulse durations, energy transfer from photons to plasmons requires ~11 fs in the present metamaterial. This value corresponds to a frequency limit of ~90 THz, which can be increased by carefully designing the metamaterial based on the Lorentz oscillator model. The optical switches made of these metamaterials are promising for future generation of optical signal processing and ultrafast optical data transfer in coherent optical networks.

**Acknowledgments**

V N, S H and N.I.Z. would like to thank the Singapore Ministry of Education Academic Research Fund Tier 3 (Grant MOE2011-T3-1-005 and MOE2016-T3-1-006) for financial support. J V would like to thank the support from the Defence Science and Technology Laboratory (Grant DSTLX1000064081).

**References**

1. Atwater H A, Polman A, Plasmonics for improved photovoltaic devices, *Nat Mater*, 9(2010)205–213.

2. Aydin K, Ferry V E, Briggs R M, Atwater H A, Broadband polarization-independent resonant light absorption using ultrathin plasmonic super absorbers, *Nat Commun*, 2(2011)517; doi.org/10.1038/ncomms1528.

3. B G Bagley, Chapter 7 - Materials, Properties, and Choices A2 - in Optical Fiber Telecommunications, eds: Stewart M, Chynoweth A G, (Academic Press), 1979, pp 167–231.

4. Chong Y D, Ge L, Cao H, Stone A D, Coherent Perfect Absorbers: Time-Reversed Lasers, *Phys Rev Lett*, 105(2010)053901; doi.org/10.1103/PhysRevLett.105.053901.

5. Zhang J, MacDonald K F, Zheludev N I, Controlling light-with-light without nonlinearity, *Light Sci Appl*, 1(2012) e18; doi.org/10.1038/lsa.2012.18

6. Fang X, Tseng M L, Ou J-Y, MacDonald K F, Tsai D P, Zheludev N I, Ultrafast all-optical switching via coherent modulation of metamaterial absorption, *Appl Phys Lett*, 104(2014)141102; doi.org/10.1063/1.4870635.

7. Roger T, Vezzoli S, Bolduc E, Valente J, Heitz J J F, Jeffers J, Soci C, Leach J, Couteau C, Nikolay I. Zheludev N I, Faccio D, Coherent perfect absorption in deeply subwavelength films in the single-photon regime, *Nat Commun*, 6(2015)7031; doi.org/10.1038/ncomms8031.

8. Liu N, Langguth L, Weiss T, Kästel J, Fleischhauer M, Pfau T, Giessen H, Plasmonic analogue of electromagnetically induced transparency at the Drude damping limit, *Nat Mater*, 8(2009)758–762.

9. Nie G, Shi Q, Zhu Z, Shi J, Selective coherent perfect absorption in metamaterials, *Appl Phys Lett*, 105(2014)201909; doi.org/10.1063/1.4902330.

10. Linic S, Christopher P, Ingram D B, Plasmonic-metal nanostructures for efficient conversion of solar to chemical energy, *Nat Mater*, 10(2011)911–921.

11. Lee M, Kim J U, Lee J S, Lee B I, Shin J, Park C B, Mussel-Inspired Plasmonic Nanohybrids for Light Harvesting, *Adv Mater*, 26(2014)4463–4468.

12. Fangli L, Y D Chong, Adam S, Polini M, Gate-tunable coherent perfect absorption of terahertz radiation in graphene. *2D Mater*, 1(2014) 031001; doi.org/10.1088/2053-1583/1/3/031001.

13. Papaioannou M, Plum E, Valente J, Rogers E T F, Zheludev N I, Two-dimensional control of light with light on metasurfaces, *Light Sci Appl*, 5(2016) e16070; doi.org/10.1038/lsa.2016.70.

14. Fang X, MacDonald K F, Plum E, Zheludev N I, Coherent control of light-matter interactions in polarization standing waves. *Sci Rep*, 6(2016)31141; doi.org/10.1038/srep31141.

15. Kita K, Nozaki K, Takata K, Shinya A, Notomi M, Ultrashort low-loss Ψ gates for linear optical logic on Si photonics platform, *Commun Phys*, 3(2020)33; doi.org/10.1038/s42005-020-0298-2.

16. Nalla V, Valente J, Sun H, Zheludev N I, 11-fs dark pulses generated via coherent absorption in plasmonic metamaterial, *Opt Express*, 25(2017)22620–22625.

17. Kim D S, Hohng S C, Malyarchuk V, Yoon Y C, Ahn Y H, Yee K J, Park J W, Kim J, Park Q H, Lienau C, Microscopic Origin of Surface-Plasmon Radiation in Plasmonic Band-Gap Nanostructures, *Phys Rev Lett*, 91(2003)143901; doi.org/10.1103/PhysRevLett.91.143901.

18. Zentgraf T, Zentgraf T, Christ A, Kuhl J, Giessen H, Tailoring the Ultrafast Dephasing of Quasiparticles in Metallic Photonic Crystals. *Phys Rev Lett*, 93(2004)243901; doi.org/10.1103/PhysRevLett.93.243901.

19. Shcherbakov M R, Vabishchevich P P, Shorokhov A S, Chong K E, Choi D-Y, Staude I, Andrey E. Miroshnichenko A E, Neshev D N, Fedyanin A A, Kivshar Y S, Ultrafast All-Optical Switching with Magnetic Resonances in Nonlinear Dielectric Nanostructures. *Nano Lett*, 15(2015)6985–6990.

20. Fang X, MacDonald K F, Zheludev N I,, Controlling light with light using coherent metadevices: all-optical transistor, summator and invertor, *Light Sci Appl*, 4(2015)e292; doi.org/10.1038/lsa.2015.65.

21. Birr T, Zywietz U, Chhantyal P, Chichkov B N, Reinhardt C, Ultrafast surface plasmon-polariton logic gates and half-adder, *Opt Express*, 23(2015)31755-31765.

22. Rao S M, Heitz J J F, Roger T, Westerberg N, Faccio D, Coherent control of light interaction with graphene, *Opt Lett*, 39(2014)5345–5347.

23. Zhang J, Guo C, Liu K, Zhu Z, Ye W, Yuan X, Qin S, Coherent perfect absorption and transparency in a nanostructured graphene film, *Opt Express*, 22(2014)12524–12532.

24. An Hui, L., H.K. Tsang, and L.Y. Chan, Two new transform-limited criteria and their applications in ultrafast optics and soliton systems, *IEEE J Quant Electron*, 32(1996)2064–2070.

25. Johnson P B, Christy R W, Optical Constants of the Noble Metals, *Phys Rev B*, 6(1972)4370–4379.

26. Olmon R L, Slovick B, Johnson T W, Shelton D, Oh S-H, Boreman G D, Raschke M B, Optical dielectric function of gold, *Phys Rev B*, 86(2012)235147; doi.org/10.1103/PhysRevB.86.235147.

27. Sönnichsen C, Plasmons in metal nanostructures, Ph D Thesis, Cuvillier Verlag: Göttingen, Germany: Ludwig-Maximilians-University of Munich, 2001.

28. Fedotov V A, Rose M, Prosvirnin S L, Papasimakis N, Zheludev N I, Sharp Trapped-Mode Resonances in Planar Metamaterials with a Broken Structural Symmetry, *Phys* Rev *Lett*, 99(2007)147401; doi.org/10.1103/PhysRevLett.99.147401.

# Supplementary Information

# Coherent Plasmonic Absorption in the Femtosecond Regime

Venkatram Nalla[1], Xu Fang[2], João Valente[2], Handong Sun[1], and Nikolay I Zheludev[1,2]

*[1]Centre for Disruptive Photonic Technologies, The Photonic Institute, Nanyang Technological University, Singapore*

*[2]Optoelectronics Research Centre and Centre for Photonic Metamaterials,*
*University of Southampton, Highfield, Southampton, UK*

## 1 Travelling-wave spectra

Figure S1 shows the travelling-wave transmission, reflection, and absorption spectra for both the metamaterial and the unstructured gold thin film, which were measured using a microspectrophotometer.

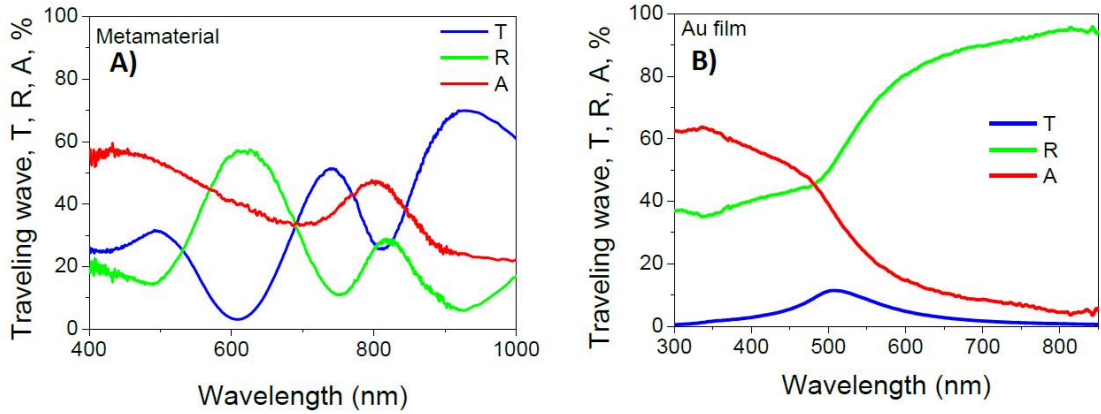

**Fig S1.** **Travelling-wave transmission, reflection, and absorption spectra** (A) the metamaterial, and (B) the unstructured gold film.

## 2 Model to simulate coherent absorption

A vanishingly thin film can be described as a 4-port device, with the input and output field amplitudes linked by a linear scattering matrix as follows:

$$\begin{pmatrix} E_\delta \\ E_\gamma \end{pmatrix} = \begin{pmatrix} s & s+1 \\ s+1 & s \end{pmatrix} \begin{pmatrix} E_\alpha \\ E_\beta \end{pmatrix} \tag{1}$$

where $s$ is the complex scattering coefficient. Here, a perfect beam splitter film that does not absorb, but transmits a travelling wave in proportion 50:50 is described by the complex scattering coefficient $s = -0.5 \pm 0.5\ i$. A perfect absorber film that can provide total absorption in a standing wave shall absorbs, reflects, and transmits travelling wave in proportion 50:25:25 is described by $s = -0.5$. A perfect, non-dissipating mirror corresponds to $s = -1$. The normalized intensities at the output ports are:

$$I_\delta/I = |s|^2 + |s+1|^2 + 2Re[(s+1)\ s^*\ e^{-i\theta}] \tag{2}$$

$$I_\gamma/I = |s|^2 + |s+1|^2 + 2Re[(s+1)\ s^*\ e^{i\theta}] \tag{3}$$

Here, we assume that incident waves are of the same amplitude $E$ and intensity $I = I_\alpha = I_\beta = |E|^2$. The mutual phase difference $\theta$ between the incident waves is introduced as $E_\alpha = Ee^{i\theta}$ and $E_\beta = E$, where $\theta = x\pi$ and $x$ is dictated by the displacement from the centre of the standing wave. For instance, $\theta = 0$ and $\theta = \pi$ correspond to a node and an antinode of the standing wave, respectively.

Equations (2) and (3) implies that, for the ideal thin-film absorber ($s = -0.5$), $I_\gamma$ and $I_\delta$ oscillate in phase upon moving the film in the standing wave: $I_{\gamma,\delta}/I = 0.5\ (1 - \cos\theta)$. For the non-absorbing beam-splitter ($s = -0.5 - 0.5i$), the output intensities oscillate out of phase: $I_{\delta,\gamma}/I = 1 \pm \sin\theta$. Obviously, for the case of perfect mirror, ($s = -1$), intensities of the reflected waves do not depend on the mirror position $I_{\delta,\gamma}/I = 1$.

## 3 Further analysis on the Lorentz oscillator model

The main text analyzes the plasmonic coherent absorption in the metamaterial by using a simple Lorentz oscillator model. The metamaterial has a maximal operation speed of ~90 THz. Further analysis using the model indicates that, by engineering the absorption of the metamaterial, even higher speed can be achieved (Fig S2).

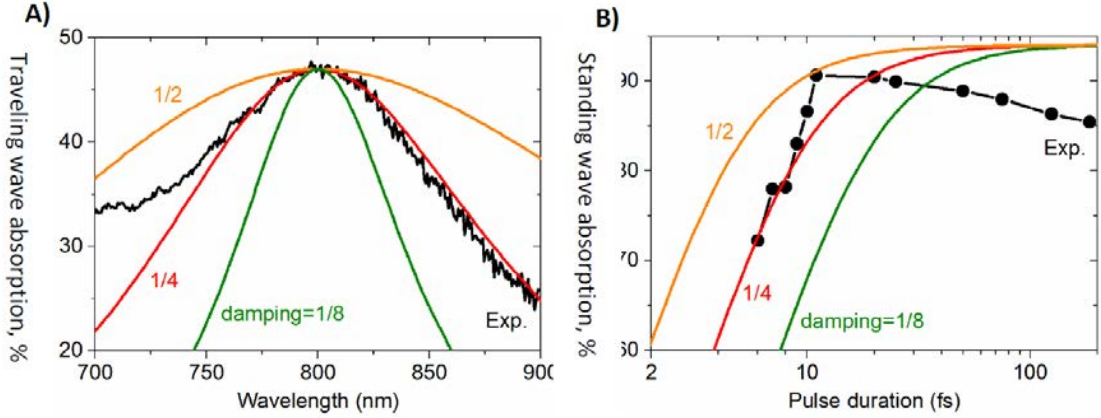

Fig S2. **Absorption calculated by using the Lorentz oscillator model**. The metamaterial absorber can be engineered to achieve high standing-wave absorption at very short pulse duration. Metamaterials with different damping show A) different travelling-wave absorption and B) different standing-wave absorption.

Figure S2 shows three metamaterials absorption simulated using the same Lorentz oscillator model. The resonance wavelength and maximal absorption are the same for all three cases, while the damping is $\psi = \omega_0/2$, $\psi = \omega_0/4$ (for the samples fabricated and investigated in the main manuscript), $\psi = \omega_0/8$. It is clear that by engineering the plasmon relaxation time in the metamaterial, we can control the time scale of the optical transient dynamics in the ultrafast modulation. A highly damped oscillator ($\psi = \omega_0/2$), which gives a broad peak in the travelling-wave absorption, allows for high standing-wave absorption for very short pulses.

## 4 Beyond the Lorentz oscillator model

In the experiment, we observed on the metamaterial sample that the standing-wave absorption slightly decreases with increasing pulse duration beyond 11 fs. This behavior cannot be explained by using the linear Lorentz model, and is attributed to the Landau damping and saturable absorption. Although this effect is very weak in the results given in the main text, here we show that it can be relatively strong at other excitation conditions. Figure S3 shows the pulse duration dependent travelling-wave and standing-wave absorption for different laser fluences. The decrease in the absorption for pulses longer than 11 fs is seen in various cases. It could be attributed to the Landau damping [4] or saturable absorption [5,6]. It is well known that, gold nanoparticles/structures are good nonlinear saturable absorbers at plasmon resonance due to Pauli blocking. Saturable absorption in gold nanostructures could be more effective for longer pulses due to the thermalization of electrons [7,8].

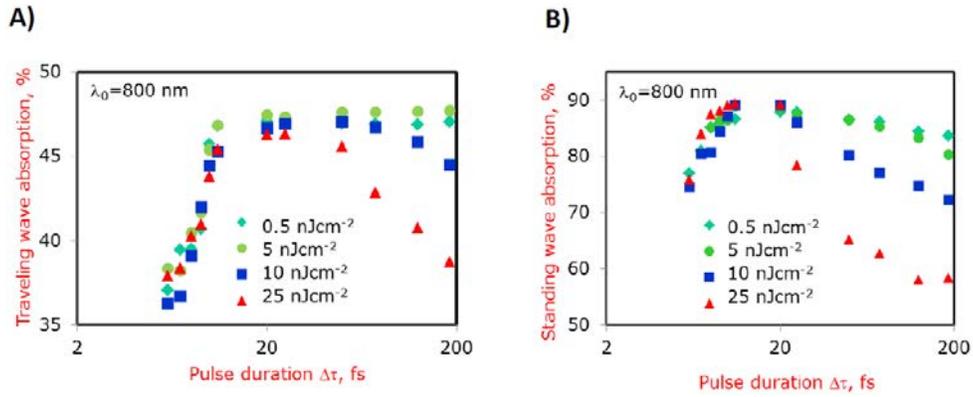

Fig S3. Pulse duration dependent absorption A) travelling-wave absorption and B) standing-wave absorption for the metamaterial at various laser fluences.

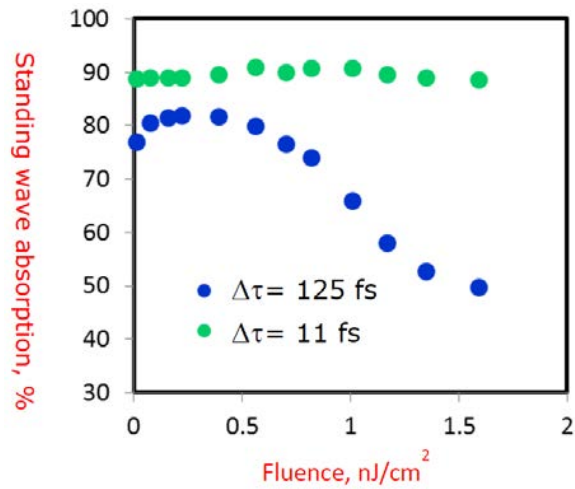

Fig S4. Nonlinear standing wave absorption. Fluence dependent standing-wave absorption for the metamaterial sample at two different pulse durations.

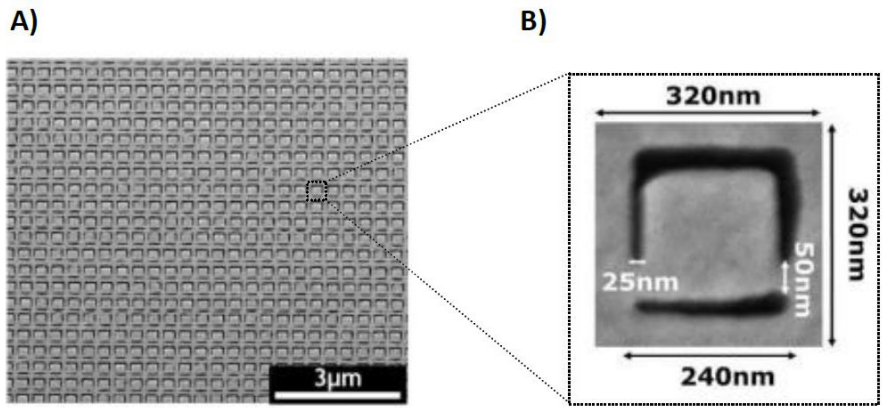

Fig S5. SEM imgages of the metamaterial. (A) Part of the whole array. (B) A single unit cell.

Finally, we also studied the fluence dependent standing-wave absorption at two different pulse durations. As the absorption should be a constant in the absence of any nonlinear effect, its variation with the fluence directly reveals the magnitude of the nonlinear effects. In this regard, the long pulse of 125 fs induces much stronger nonlinearity than the short pulse of 6 fs.

**Referecnes**

1. Chong Y D, Ge L, Cao H, Stone A D, Coherent Perfect Absorbers: Time-Reversed Lasers. *Phys Rev Lett*, 105(2010)053901; doi.org/10.1103/PhysRevLett.105.053901.

2. Thongrattanasiri S, Koppens F H L, de Abajo F J G, Complete Optical Absorption in Periodically Patterned Graphene. *Phys Rev Lett*, 108(2012)047401; doi.org/10.1103/PhysRevLett.108.047401.

3. Hägglund C, Apell S P, Kasemo B, Maximized Optical Absorption in Ultrathin Films and Its Application to Plasmon-Based Two-Dimensional Photovoltaics, *Nano Lett*, 10(2010)3135–3141.

4. Brongersma M L, Halas N J, Nordlander P, *Nat Nanotechno*, 10(2015)25–34.

5. Venkatram N, Kumar R S S, Rao D N, Medda S K, De S, De G, *J Nanosci Nanotechnol*, 6(2006)1990–1994.

6. Ren M, Plum E, Xu J, Zheludev N, *Nat Commun*, 3(2012)833; doi.org/10.1038/ncomms1805.

7. Rotenberg N, Bristow A, Pfeiffer M, Betz M, van Driel H, *Phys Rev B*, 75(2007)155426; doi.org/10.1103/PhysRevB.75.155426.

8. Boyd R W, Shi Z, Leon I D, *Opt Commun*, 326(2014)74-79.